# Dynamic Breaking of Mirror Symmetry in Spin-Dependent Electron Transport through Chiral Media Causes Enantiomeric Excesses

Yossi Paltiel,*[1] Daniel Goldberg,[1] Nir Yuran,[1] Shira Yochelis,[1] Jia Hao Soh,[2] Christopher Seibel,[2,3] Jurgen Gauss,[4] Shmuel Zilberg,[5] S. Furkan Ozturk,*[6] Jonas Fransson,*[7] Anna I. Krylov,*[2] and Ron Naaman*[8]

1) Applied Physics Institute, Center for Nanoscience and Nanotechnology, Hebrew University of Jerusalem, Jerusalem 91904, Israel
2) Department of Chemistry, University of Southern California, Los Angeles, California 90089, United States
3) Department of Physics and Research Center OPTIMAS, RPTU Kaiserslautern-Landau, 67663 Kaiserslautern, Germany
4) Department Chemie, Johannes Gutenberg-Universität Mainz, 55128 Mainz, Germany
5) Department of Chemical Sciences, Ariel University, Ariel 4076414, Israel
6) Division of Geological and Planetary Sciences, California Institute of Technology, Pasadena, CA 91125, United States
7) Department of Physics and Astronomy, Uppsala University, Box 516, 752 37 Uppsala, Sweden
8) Department of Chemical and Biological Physics, Weizmann Institute, Rehovot 7610001, Israel

Y. Paltiel Email: paltiel@mail.huji.ac.il
S. F. Ozturk Email: ozturk@caltech.edu
J. Fransson Email: jonas.fransson@physics.uu.se
A. I. Krylov Email: krylov@usc.edu
R. Naaman Email: ron.naaman@weizmann.ac.il

Teaser: Spin dependent electron transport through chiral system is enantiospecific and may provide explanation for homochirality in life.

**Abstract**

Two fundamental questions have puzzled scientists for more than 150 years. "How did life become homochiral?"  and "why was this specific handedness selected?"

Recently, it has been shown that homochirality could have emerged through the enantioselective interactions of molecules with magnetic substrates due to the asymmetric crystallization of an RNA precursor on a magnetite substrate, abundant on early Earth. This phenomenon is based on the chirality-induced spin selectivity (CISS) effect. Despite its robustness, this model could not provide an answer to the second question: why one specific handedness (D for RNA) was selected.

Here we demonstrate that spin-involving processes can have different outcomes in the two enantiomers of chiral molecules. In chiral molecules with unpaired electrons or while electrons are passing through them, the total angular momentum vector, **J,** is aligned along the "easy axis", which is defined by the magnetic anisotropy induced by the spin-orbit coupling and asymmetry of the molecular field. The magnitude **J** is the same for both enantiomers, but the vectors may be aligned differently relative to the molecular frame in the two enantiomers. This difference can be quantified by, for example, by the angle between **J** and electric dipole moment of the molecule, **μ**. We show by direct measurements, theory, and *ab initio* calculations that dynamic spin processes in chiral molecules could result in different efficiencies of spin-related phenomena, including the interaction of chiral molecules with magnetic surfaces. The findings may provide an explanation for the specific homochirality in nature.

## Introduction

It is commonly accepted that besides the slight effect of the weak force, the energetics of two enantiomers is identical. (*1*) This is the background to the mystery involved in homochirality in nature. Although it is relatively easy to understand why homochirality persists in nature, to date, there has been no clue as to why in life this handedness exists. Specifically, why is it that almost all the amino acids in living organisms have L symmetry, namely, they turn linearly polarized light to the left, whereas sugars and nucleic acids are of the D type, namely, they rotate linearly polarized light to the right. Over the years, several schemes have been proposed regarding the emergence of chirality in life, but none of them have explained a specific handedness.(*2*) Based on the identical energetics of the two enantiomers and symmetry-imposed mapping of their properties, it has been generally expected that the absolute value of any physical effect will be the same for the two enantiomers.(*3*)

The chirality-induced spin selectivity (CISS) effect was discovered about two decades ago.(*4*) According to this effect, electrons' motion through a chiral system depends on their spin. For one handedness the electrons' spin is aligned preferentially parallel to their velocity, whereas for the other handedness it is antiparallel to the velocity. Naturally, it was expected that all the results related to the CISS effect would be perfectly symmetric for the two enantiomers. For example, the conductance of β-spin through the left-handed molecule will be the same as the conductance of α-spin through the right-handed molecule. However, in numerous cases, the results for the two enantiomers were indeed with opposite signs, but they were not quite symmetrical. This was attributed by us and by others to the impurity of the samples. Moreover, many recent detailed quantitative studies clearly demonstrated that the observed asymmetry is real.(*5*) It must be stated upfront that the results obtained relate to the electrons' dynamics in those molecules that change the kinetics of processes and that the asymmetry is not related to an equilibrium property. It is also important that all the relevant experiments either use magnetic substates with a specific magnetization direction or were sensitive to a specific spin state.

Here we present the physical process that may explain the asymmetry, followed by experimental results that demonstrate the asymmetry in CISS experiments. We then show results of *ab initio* calculations that confirm the asymmetry between the spin-polarization in the two enantiomers. Finally, we discuss the relationship of asymmetry to the origin of the specific homochirality in life.

## Results and Discussion

*How the loss of symmetry in the CISS experiments can be explained?*

The spin-polarization observed in CISS-related experiments depends on the projection of the total angular momentum **J** on the path of the transmitted electrons. This projection in turn depends on the SOC operator. Hence, we begin with a phenomenological discussion focused on these properties.

The “enantiomeric dependence” of the SOC was recognized before.(*6-. 11)* Hence, as will be shown below (eq.4), the SOC term, $\mathbf{V}_{\text{SOC}}$ , of a chiral system, can be presented schematically as including two components:

$$\mathbf{V}_{\text{SOC}} = \mathbf{V}_{\text{SOC}}^{\text{a}} + \mathbf{V}_{\text{SOC}}^{\text{chiral}} \qquad (1)$$

The first term, $\mathbf{V}_{\text{SOC}}^{\text{a}}$, arises from the atomic SOC and its sign depends on the shell filling the atoms (the local term); however, it is not related to the chiral structure. The second contribution, $\mathbf{V}_{\text{SOC}}^{\text{chiral}}$, arises from the topology; the sign of some of its components depends on the handedness (the non-local term).

Equation (1) can be easily verified by the following consideration: The starting point is the general expression for the spin-orbit coupling, that is,

$$\mathbf{V}_{\text{soc}} = \xi[\boldsymbol{E} \times \boldsymbol{p}] \cdot \boldsymbol{\sigma} \qquad (2)$$

where $\xi = 1/4c^2$ in atomic units, c being the speed of light, $\boldsymbol{E}$ is the effective electric field, $\boldsymbol{p}$ is the momentum operator, and $\boldsymbol{\sigma}$ is the vector of Pauli matrices. As is evident from eq. 2 in the SI, the “atomic” contribution does not depend on the structure, since the spatial extension of the electric field outside the atomic vicinity makes a negligible difference to the intra-site coupling. This contribution is, moreover, element-specific and can be considered constant with respect to the handedness of the structure.

The second term in Eq. (1), however, is non-local, since it captures the overlap of the orbitals at different sites via the SOC potential. Consequently, it strongly depends on the structure and may acquire different phases, depending on the system’s structure. The easiest way to understand this feature is by Fourier transforming the state functions:

$\phi_m(\boldsymbol{r}) = \int \phi_{\boldsymbol{k}} e^{i\boldsymbol{k}\cdot(\boldsymbol{r}-\boldsymbol{r}_m)} d\boldsymbol{k}/\Omega$, where $\Omega$ is the pertinent volume.

Here, as an example, we consider a helical structure with $\mathbb{M}$ sites, the sites are distributed according to $\boldsymbol{r}_m = (a\cos\varphi_m, \pm a\sin\varphi_m, c\varphi_m/2\pi)$, where $a$ is the radius, $c$ is the length, $\varphi_m = 2\pi(m-1)/(\mathbb{M}-1)$, and $1 \leq m \leq \mathbb{M}$ for the counterclockwise (+) and clockwise (-) helix. Hence, the differences between two nearest neighboring sites in the two enantiomers are equal to, e.g.,

$$\boldsymbol{r}_m - \boldsymbol{r}_{m+1} = (a\mathcal{R}\cos(\varphi_m - \eta), \pm a\mathcal{R}\sin(\varphi_m - \eta), -c/(\mathbb{M}-1)), \qquad (3)$$

where $\mathcal{R} = \sqrt{2(1-\cos^2\varphi_0)}$, $\tan\eta = \cot(\varphi_0/2)$, and $\varphi_0 = 2\pi/(\mathbb{M}-1)$. Therefore, the phase contribution to the SOC is given by $\boldsymbol{k}\cdot(\boldsymbol{r}_m - \boldsymbol{r}_{m+1})$ and it is enantiospecific.

In terms of the discussion above, the SOC can finally be written as

$$\mathbf{V}_{\text{SOC}} = \sum_{mn} \psi_m^\dagger(\delta_{mn}\boldsymbol{L}_m + \boldsymbol{K}_{mn})\cdot\boldsymbol{\sigma}\psi_n,$$

$$\boldsymbol{L}_m = \xi\int \phi_m^*(\boldsymbol{E}\times\boldsymbol{p})\phi_m d\boldsymbol{r},$$

$$\boldsymbol{K}_{mn} = \xi\int \phi_{\boldsymbol{k}+\boldsymbol{q}}^* V_{\boldsymbol{q}}\phi_{\boldsymbol{k}}(\boldsymbol{k}\times\boldsymbol{q})e^{-i\boldsymbol{q}\cdot\boldsymbol{r}_m - i\boldsymbol{k}\cdot(\boldsymbol{r}_m-\boldsymbol{r}_n)}\frac{d\boldsymbol{k}}{\Omega}\frac{d\boldsymbol{q}}{\Omega}. \quad (4)$$

where $V_{\boldsymbol{q}}$ is the Fourier transform of the electron confinement potential $V(\mathbf{r})$ which is connected to the electric field via $\mathbf{E}(\mathbf{r}) = -\nabla V(\mathbf{r})$. Orbitals $\phi_m$ can always be chosen to be real valued. Considering the nearest neighbors as in the above, the imaginary part of the SOC is

$$-i\xi\int \phi_{\boldsymbol{k}+\boldsymbol{q}}V_{\boldsymbol{q}}\phi_{\boldsymbol{k}}(\boldsymbol{k}\times\boldsymbol{q})\sin A_{\boldsymbol{kq}m}\cos B_{\boldsymbol{k}m}\frac{d\boldsymbol{q}}{\Omega}\frac{d\boldsymbol{k}}{\Omega}$$

$$\mp i\xi\int \phi_{\boldsymbol{k}+\boldsymbol{q}}V_{\boldsymbol{q}}\phi_{\boldsymbol{k}}(\boldsymbol{k}\times\boldsymbol{q})\cos A_{\boldsymbol{kq}m}\sin B_{\boldsymbol{k}m}\frac{d\boldsymbol{q}}{\Omega}\frac{d\boldsymbol{k}}{\Omega}, \quad (5)$$

$$A_{\boldsymbol{kq}m} = \boldsymbol{q}\cdot\boldsymbol{r}_m + k_x a\mathcal{R}\cos(\varphi_m - \eta) - k_z c/(\mathbb{M}-1),$$

$$B_{\boldsymbol{k}m} = k_y a\mathcal{R}\sin(\varphi_m - \eta).$$

The outcome from the above is that although the real part of the SOC is the same for both enantiomers, the sign of the imaginary part is different and may give rise to a different orientation of **J** relative to the molecular frame in the two enantiomers.

From these expressions, we can discern a few aspects that are important in the context of chiral structures. The first integral in Eq. (5) acquires the same sign regardless of the helicity, since $A_{\boldsymbol{kq}m}$ is the same for the two enantiomers. The second integral, on the other hand, has opposite signs ($\mp$) for the two enantiomers, which either reduce or enhance the contribution from the first integral. Furthermore, the second term requires that the potential $V_{\boldsymbol{q}}$ has no inversion symmetry such that $V_{-\boldsymbol{q}} \neq -V_{\boldsymbol{q}}$, a requirement which, by definition, is fulfilled for chiral structures. This is a clear indication that the geometrical contribution to the SOC has deep significance for chiral structures, which is not present for achiral ones.

Since, as shown in Eq. (3), in chiral systems, the phase of the SOC is enantiospecific, the total angular momentum vector, **J**, may be aligned at a different angle in each enantiomer, relative to the molecular frame. Consequently, there is a different projection, $\mathbf{J}_z$, of the angular momentum on a specific path of an electron for each of the enantiomers. Since **J=L+S**, when **L** and **S** are the orbital angular momentum and the spin, respectively, the values of the projection of **S**, on a specific axis, $\mathbf{S}_z$, is also different and it can be represented by combining the two spin states, α and β (see Figure 1). For each enantiomer this combination can be different and not necessarily be symmetric, namely, one cannot assume that the spin combination is simply aα+bβ for one enantiomer and bα+aβ for the other, where a and b are two coefficients. The spin-polarization measured by a spin dependent detector, like magnet for example, for electrons conducted along the *z*-axis, can be defined as $P = \frac{I_\alpha - I_\beta}{I_\alpha + I_\beta}$ and $I_\alpha, I_\beta$ are the spin currents of the two different spins. Hence for the two enantiomers, one can obtain different absolute value of the spin-polarization.

Therefore, when electron is conducted through a closed-shell, the negative ion states must be considered for the angular momentum analysis.

This phenomenon was observed in previous calculations; where, it was clearly stated that although the energies of the states in the two enantiomers are identical, the ratio between the two spins varies and it does not reflect the perfect "mirror image symmetry" expected.(*7,8,9)*

The description given above can be formulated in terms of a pseudoscalar, a quantity that can distinguish two enantiomers. In the case of a transmitted electron, the relevant quantities are scalar product of $\mathbf{S}\cdot\boldsymbol{\mu}$ and $\mathbf{S}\cdot\mathbf{p}$, where **μ** and **p** are the electric dipole moment of the molecule and the linear momentum of the electron, respectively. Because **S** is an axial vector, and **μ** and **p** are polar vectors, their product changes sign upon inversion, which makes them pseudoscalars. The important distinction is that $\mathbf{S}\cdot\boldsymbol{\mu}$ is a time-odd pseudoscalar and $\mathbf{S}\cdot\mathbf{p}$ is a time-even pseudoscalar. Following Barron's argumentation,(*3)* one can obtain different outcomes for the enantiomers for processes expressed as time-even properties pseudoscalars. Hence, different values of a static property $\mathbf{S}\cdot\boldsymbol{\mu}$ set the stage for different outcomes that can be realized in dynamic systems, when the transmitted electron has linear momentum **p**.

An important note about the CISS: In the above discussion we related to the SOC of the system in its adiabatic electronic states, implicitly assuming the Born-Oppenheimer approximation. However, when an electron passes through the system, one must consider the coupling of electronic and nuclear degrees of freedom, which can be understood in terms of electronic polarization and vibrational excitations. The ensuing nonadiabatic transitions result in the transfer of electronic energy into nuclear vibrations. Model simulations of Fay and Limmer (***Error! Bookmark not defined.****)* have shown that such relaxation of an electronically excited state can convert transversal spin-polarization into different spin populations.

Multiple experimental observations hint that nuclear vibrations indeed play an important role in CISS. In addition to these non-Born-Oppenheimer effects,(*12-14)* nuclear vibrations can also contribute to enhancing the magnitude of the effective SOC via vibronic SOC effects.(*15*)

*Asymmetry in CISS experiments*

The asymmetry between the two enantiomers was realized in numerous experiments. However, we decided to present here only one type of measurements, namely, anomalous Hall studies of several interesting systems. In the first experiment we measured the Hall signal in chiral gold (*5)* and chiral silver. We prepared chiral gold and silver film in a process based on electrodeposition of the metals from an electrolyte containing the metal salts ($Na_3Au(S_2O_3)_2$), $AgNO_3$, and either L or D tartaric acid.(*16,17)*

Figure 2 presents the Hall signal obtained with the chiral gold. No external magnetic field was applied. The Hall signal is presented as a function of the voltage applied between the source and drain electrodes; the plots for the L and D gold have opposite signs, as expected. The slope value for the L gold is larger than that for the D gold. For achiral gold, no signal is observed. Although the tartaric acid purity is 99.5 and 99% for the L and D enantiomers, the asymmetry between the slopes is about 30% when the asymmetry is defined as $A = \frac{S_L - S_D}{S_L + S_D}$ when $S_L$ and $S_D$ are the slopes for the L and D enantiomers, respectively. For the chiral silver the asymmetry measured is about 10%.

In another class of experiments the chiral molecules were adsorbed on top of a Hall device and the changes in the Hall signal were measured for the two enantiomers on both the Hall configuration and the Van der Pauw configuration to remove any changes arising from the contacts asymmetry.(18) Table 1 summarizes all the results measuring the Hall slope for four systems each measured with 6 different devices.

Chiral gold and chiral silver were chosen to present a similar system but with different magnitude of the atomic SOC. Adding an oxide layer was studied to reduce coupling with the magnet's surface. The results obtained for chiral gold and chiral silver indicate that although in both cases significant asymmetry is observed, the asymmetry is larger with gold. We ascribe this change to the larger magnitude of SOC of Au, which may enhance the asymmetry. Regarding the polyalanine molecules adsorbed on the Hall device, the SOC of gold, compared to silver, in the hybrid metal-chiral monolayer system controls the asymmetry. By adding an oxide layer between the chiral molecules and the metal, the coupling between them is reduced; hence, one expects that the metal will contribute less to an effective SOC; consequently, the asymmetry is reduced.

*Ab initio calculations and a model example*

A recent study introduced a model molecular platform for emulating the CISS effect.(***Error! Bookmark not defined.****,19)* The design, inspired by the molecules used for optical cycling,(*20*) entails a chiral scaffold with one or two SrO- groups attached to it (one of such molecules is shown in Fig. 3). A special feature of SrO- groups (or other group II metals) is that they support localized atomic-like states, with one unpaired electron occupying *s*-, *p*-, and *d*-like orbitals.

We describe the electronic structure of this molecule using a state-interaction approach. The non-relativistic electronic states were computed using equation-of-motion coupled-cluster method for electron attachment with single and double excitations (EOM-EA-CCSD)(*21,22)* and were used to construct the state-interaction Hamiltonian, which, depending on the calculation, may include the following terms: SOC terms computed as matrix elements of the full-electron Breit–Pauli Hamiltonian, the Zeeman term (to account for an applied magnetic field), and time-dependent electric field perturbation (to simulate electronic excitations). Details of the calculations can be found in Refs. 8 and 19, and in the Supporting Information.

In these calculations the SOC is described by the spin-orbit part of the Breit-Pauli Hamiltonian: (*23*)

$$H_{SO}^{BP} = \frac{\hbar e^2}{2m_e^2 c^2}\left[\sum_i \boldsymbol{h}^{SO}(i)\cdot \boldsymbol{s}(i) - \sum_{i\neq j} \boldsymbol{h}^{SOO}(i,j)\cdot\left(\boldsymbol{s}(i) + 2\boldsymbol{s}(j)\right)\right], \qquad (6)$$

$$\boldsymbol{h}^{SO}(i) = \sum_K \frac{Z_K(\boldsymbol{r}_i - \boldsymbol{R}_K)\times \boldsymbol{p}_i}{|\boldsymbol{r}_i - \boldsymbol{R}_K|^3} = \sum_K \frac{Z_K}{r_{iK}^3}(\boldsymbol{r}_{iK}\times \boldsymbol{p}_i), \qquad (7)$$

$$\boldsymbol{h}^{SOO}(i,j) = \frac{(\boldsymbol{r}_i - \boldsymbol{r}_j)\times \boldsymbol{p}_i}{|\boldsymbol{r}_i - \boldsymbol{r}_j|^3} = \sum_{ij}\frac{1}{r_{ij}^3}\left(\boldsymbol{r}_{ij}\times \boldsymbol{p}_i\right), \qquad (8)$$

where $\mathbf{r}_i$, $\mathbf{p}_i$ and $\mathbf{s}$(i) are the coordinate, momentum, and spin operators of the $i^{th}$ electron, and $R_K$ and $Z_K$ denote the coordinates and the charge of the $K^{th}$ nucleus. Both one- and two-electron terms are important and are included in the *ab initio* calculations reported here,(*24*) but for the purpose of symmetry analysis it is sufficient to focus on the one-electron term, $\mathbf{h}^{SO}(i)$. This operator includes cross product of **r** and **p** and, therefore, has handedness (just like angular momentum). As one can see, Eq. (7) includes summation over all nuclei and, therefore, one may expect that $\mathbf{h}^{SO}$ may sense molecular structure including chiral environment. On the other hand, due to denominator, SOC is a local operator. Moreover, in molecules such as those used here to illustrate spin-polarization,(*8,22)* the electronic states themselves are localized quite far from the chiral center. As we illustrate by numeric example (in the SI), molecular environment affects the shape of the orbitals (for example, 3 atomic *p*-orbitals), and the SOC operator (or, specifically, angular momentum part of it) senses the shape of the orbitals and reports on it. In particular, in chiral systems, molecular field imparts differently handed shapes of the orbitals in the enantiomers, which then result in different phases of the matrix elements of the SOC operator due to the intrinsic handedness of $\mathbf{h}^{SO}$. In other words, chirality of the molecular environment manifests in the shapes of the molecular orbitals, even when they are localized far from the chiral centers, and these shapes can be probed and reported by the SOC matrix elements. As we show by a numeric example, different phases of SOC result in different spin-polarization in the enantiomers.

Indeed, the calculations have demonstrated the emergence of an enantiomer-specific transversal spin-polarization that can be induced by photoexcitation and probed by applying a magnetic field, akin to EPR-type experiments. Spin-polarization also manifests in static calculations. Figure 3 shows the direction of spin vector **S** for one of the excited states in the two enantiomers of the model molecule. The energies of this excited state are the same for the two enantiomers and the respective charge distributions are mirror images of each other, but the spin vectors are not. Therefore, the orientation of **J** relative to the molecular frame differs in the *S* and *R* enantiomers, and the **S** vector follows **J** . The alignment can be quantified by the angle between **S** (or **J**) and **μ** where **μ** is the dipole moment of the molecule. In this example (using the dipole moment of the reference state), the angle is 107.5° and 72.5° for the *R* and *S* enantiomers, respectively (note that in these calculations, the special axis is the *x*-axis, which can be compared to the *z*-axis in Eq. (5) and Figure 1).

The authors related this spin-polarization to the different phases of the matrix elements of the Breit–Pauli SOC.(*8)* They noted that in chiral molecules the phase of the SOC differs in the *S* and *R* enantiomers. Namely, the signs of the imaginary relative to the real parts of the SOC are flipped in the two enantiomers. However, the phase alone is not sufficient, an important prerequisite for magnetic anisotropy is the non-zero angular momentum. Hence, not all electronic states of the model molecule show spin-polarization.

In addition to this example, in the SI we show calculations on even a simpler model system, an F atom surrounded by 4 He atoms in a chiral arrangement (distorted tetrahedral). The electronic states of F are 3 *p*-orbitals distorted by the cluster environment. The calculations show that weak interactions with this chiral environment result in orbital hybridization, which then leads to different phases in SOC and spin-polarization. For a detailed discussion of the *ab initio* calculations, see the SI. In addition, in the SI we present the calculations on ribose-aminooxazoline (RAO), which also illustrate different phases of the SOC for the two enantiomers. These results are relevant to the discussion presented below.

There are many indications that the rate of processes that involve chiral molecules are spin polarization dependent and they scale with the extent of spin polarization. This is the case in reactions,(25,26) in adsorption on ferromagnetic surfaces,(*27*) and in electron-transfer rates. Hence, the effect described here means that in the case of competition between two enantiomers, in a given process, the results may be significantly different for the two enantiomers, depending on the molecules and the specific path of the electrons involved in the process. As shown in Fig. 2 the Hall signal for L and D oligopeptides of polyalanine, can differ by almost a factor of three. The effect presented here indeed calls for detailed studies with different systems.

### *Asymmetry in SOC and its Relationship to Life's Homochirality*

A recent work by Ozturk and Sasselov proposed that ribose-aminooxazoline (RAO) could act as a central precursor to RNA, and that the molecular homochirality of an early genome could emerge through its interaction with magnetite ($Fe_3O_4$) surfaces via the CISS effect.(*28*) RAO has also been shown to induce a net magnetization on magnetic surfaces that were previously demagnetized, suggesting a self-reinforcing feedback mechanism between chiral molecules and magnetized surfaces.(*29*) Furthermore, the homochirality established at the level of RAO can propagate first to the genome, then to other biomolecules—from D-RNA to L-peptides, and subsequently to homochiral metabolites—offering a plausible framework for the emergence of life's homochirality in nature.(*30*)

Before discussing farther the impact of the effect presented in this manuscript on the possible origin of chirality in life, it is important to appreciate what past studies on the CISS effect concluded regarding the interaction of chiral molecules with ferromagnetic surfaces. It was established that there is enantiospecific interaction of chiral molecules with ferromagnetic surface that depends on the direction of magnetization of the surface. Namely if the north pole of the magnetic moment of the surface is pointing up one enantiomer is adsorbed preferentially, while if the north pole is pointing down the other enantiomer is preferred.(27) The results were discussed also in numerous works (*13,31,32)* and basically it is understood that when chiral molecules approach the magnetic surface, due to charge arrangement in the molecules, the molecules are charge polarized and the charge polarization is accompanied by transient spin polarization. Which spin is associated with which pole depends on the enantiomer. The accumulated charge and spin dipole moment, on the molecule, interacts via spin exchange interaction with the spins in the ferromagnet. Hence, the strength of this interaction depends on which spin is in access on the binding group of the molecule and is proportional to the extent of spin polarization. The spin exchange interaction, between the ferromagnet and the molecule, stabilizes the transient spin polarization on the molecule.

This process describing the interaction of chiral molecules with ferromagnetic surfaces is the bases for the model on the origin of life's homochirality, to which we relate here. Initially when the model was proposed, it was assumed that the specific handedness of biomolecules (e.g., D-RNA and L-peptides, or the opposite manifold) would be regionally determined across the scale of a planetary hemisphere with opposing magnetizations induced by Earth's dipolar magnetic field. Accordingly, it predicted that a biomolecule's specific handedness would emerge as a locally deterministic, but not a universally deterministic, outcome. However, our analysis here challenges the necessity of this assumption. Instead, we propose that CISS-driven homochirality at the RAO stage may inherently favor the selection of D-RNA and L-peptides in a universal manner. This selection could stem from an intrinsic asymmetry in spin-polarization: magnetite surfaces magnetized by D-RAO may acquire a stronger induced magnetization due to higher spin polarized induced by the chiral molecule, compared with those interacting with L-RAO. Although this revised model is promising, further experimental and theoretical/computational validation is needed to confirm whether the inherent asymmetry of CISS is strong enough for prebiotic materials (e.g., RAO on magnetite) to can indeed account for the universal emergence of life's homochirality with a specific handedness. Hence, within this proposed model, the enantiomeric dependent SOC may provide a direction for explaining the specific handedness of homochirality in nature.

Our experimental findings, supported by a theoretical analysis and *ab initio* calculations, indicate that for chiral molecules, the physical properties associated with spin-specific electron redistributions or conduction differ between the two enantiomers for a chosen direction of electron motion. This result challenges the long-standing assumption that enantiomers interact with physical chiral agents with identical absolute magnitudes. It is important to appreciate that typically the enantiomeric dependence of the SOC was not considered in former theoretical treatments of CISS (*33*-*35*) except of few cases, see ref. *6*-*11*, and hence the current work adds a new perspective.

It is important to appreciate that it is difficult to extract from the *ab initio* calculated SOC the "atomic" and "topological" contributions since the calculations are based on all-electron Hamiltonian expressed in the basis of atomic orbitals. The analytical expression given here allows to rationalize the differences observed in the two enantiomers.

The type of properties or interactions that could lead to different outcomes for the enantiomers must correspond to time-even pseudoscalars. The calculations on a model system suggest that one such possibility could be $\boldsymbol{S} \cdot \dot{\boldsymbol{\mu}}$ (spin interacting with a dipole moment operator *changing* in the course of nuclear vibrations).

Our work shows that in the case of charge motion through the two enantiomers, the enantiomers can be intrinsically distinct, in terms of the spin properties of this charge due to the CISS effect. These insights open a pathway toward a new level of control over asymmetric processes regulated by the electron spin.

The difference between the enantiomers can be traced to the different phases of the matrix elements of the SOC operator. Therefore, although there is no difference in the energy levels that can be observed, a difference in the ratio between the spin-state population for electrons moving in the chiral system can develop. This breaking of symmetry *per se* does not depend on temperature, although the magnitude of the SOC may; therefore, the two enantiomers show significant variations in the kinetics of many processes that involve electron motion, namely, all reactions and interactions with surfaces. The phenomenological treatment (presented in the first part of the manuscript) provides a complimentary perspective on how the chiral molecular field leads to enantiomer-specific differences in the phases of SOC.

The requisite ingredients for spin-polarization to emerge in a system are (i) chirality, (ii) SOC, and (iii) the non-zero angular momentum of the conducting state; the two latter conditions are equivalent to the requirement of magnetic anisotropy.(*36*) In the context of the CISS effect, chiral molecular environment creates chiral shapes of molecular orbitals, which then can be sensed by the SOC operator, resulting in different phases of the matrix elements. SOC prepares the states with different spin-polarizations, but it does not break symmetry because degeneracy is preserved, and the states of the two enantiomers are mapped into each other by parity and time-reversal symmetry. However, different spin-polarizations can induce different interactions of the enantiomers with other chiral agents that are sensitive to the spin or angular momentum, e.g., a magnetic field, species with unpaired electrons, spin-polarized free electrons, or chiral vibrations. The interactions that can be described as time-even pseudoscalars can yield different outcomes for the enantiomers.

This work provides a plausible and universal route for better understanding the origins of biomolecular homochirality and the specific handedness of chiral molecules in nature. The results presented here support a model for the involvement of magnetic surfaces and the CISS effect in the emergence of homochirality, presented by Ozturk and Sasselov. Moreover, it provides new insights on spin-dependent enantiospecific processes and devices.

**Acknowledgements**

**Competing interests**

The authors declare no competing interests.

**Funding statement**

R.N. acknowledges the historic generosity of the Harold Perlman Family, the support from the US Department of Energy Grant ER46430 and the partial support of the AFOSR Grant FA9550-21-1-0418. Y.P and R.N acknowledge the BSF transformative grant 3013006682. Y. P. acknowledge the partial support of the Israeli ministry of science MOS 0005959. AIK is supported by the US Department of Energy (DE-SC0022326). S.F.O. acknowledges support from startup funds provided by Caltech and from the William H. Hurt Scholarship Program.

**Author contributions**

YP- writing-original draft, investigation, writing-review&editing, methodology, Resources, Funding acquisition, data curation, validation, supervision, formal analysis, project administration, visualization

NY-Investigation, writing-review&editing, data curation, validation, Formal analysis, software

DG- Investigation, Resources, Data curation, validation, Formal analysis, software

SY- Conceptualization, Investigation, Writing- review&editing, Methodology, Resources, validation, supervision, project administrative

JHS- software, formal analysis

CS-Investigation, Writing-review&editing, Methodology, Data curation, validation, Formal analysis, software

JG-Conceptualization, writing-review&editing, Methodology, Validation, Formal analysis

SZ- Conceptualization, writing-review&editing, methodology, project administration

SFO-Writing original draft, conceptualization, investigation, writing-review&editing, Methodology, Validation, project administration

JF-writing-original draft, conceptualization, investigation, writing-review&editing, Methodology, Formal analysis

AK- writing-original draft, investigation, writing-review&editing, methodology, Resources, Funding acquisition, data curation, validation, supervision, formal analysis, project administration, visualization

RN-Written original draft, conceptualization, investigation, writing-review&editing, methodology, Resources, Funding acquisition, validation, supervision, formal analysis project administration, visualization.

**Data and Materials availability statement**

All data and code needed to evaluate and reproduce the results in the paper are present in the paper and/or the Supplementary Materials. This study did not generate new materials

**References**

1. M. Quack, G. Seyfang, G. Wichmann, Perspectives on parity violation in chiral molecules: theory, spectroscopic experiment and biomolecular homochirality, *Chem. Sci*., **13,** 10598-10643 (2022).

2. D.G. Blackmond, The Origin of Biological Homochirality. *Cold Spring Harb Perspect Biol*, **11**, a032540. (2019).

3. L. D. Barron, Symmetry and molecular chirality, *Chem. Soc. Rev*. **15,** 189 (1986).

4. B. Bloom, Y. Paltiel, R. Naaman, D. Waldeck, Chiral Induced Spin Selectivity, *Chem. Rev.* **124**, 1950–1991 (2024).

5. T. K. Das, O. Marelly, S. Yochelis, Y. Paltiel, R. Naaman, J. Fransson, Long-Range Spin Transport in Chiral Gold, *Adv. Mat*. 2506523 (2025).

6. J. S. Peter, S. Ostermann, S. F. Yelin, Chirality-induced emergent spin-orbit coupling in topological atomic lattices, *Phys. Rev. A* **109**, 043525 (2024)

7. T. P. Fay, Chirality-Induced Spin Coherence in Electron Transfer Reactions, *J. Phys. Chem. Lett.* **12**, 1407-1412 (2021).

8. C. Seibel, J. H. Soh, S. Zilberg, A. I. Krylov, Ultrafast Transversal CISS Effect Observed in a Chiral Photoswitching Molecule, *J. Phys. Chem. Lett.*, **16**, 8514 – 8522 (2025).

9. Fay, T. P.; Limmer, D. T. Origin of chirality induced spin selectivity in photoinduced electron transfer. *Nano Lett.*, **21**, 6696−6702 (2021).

10. M. Di Ventra, R. Gutierrez, G. Cuniberti, Chirality-Induced Spin−Orbit Coupling and Spin Selectivity, *J. Phys. Chem. A*, **129**, 9504−9510 (2025).

11. A. Shitade, E. Minamitani, Geometric spin–orbit coupling and chiralityinduced spin selectivity, *New J. Phys.* **22** 113023 (2020).

12. J. Fransson, Vibrational origin of exchange splitting and chiral-induced spin selectivity. *Phys. Rev. B*, **102**, 235416 (2020).

13. J. Fransson, Charge Redistribution and Spin Polarization Driven by Correlation Induced Electron Exchange in Chiral Molecules, *Nano Lett*., **21,** 3026-3032 (2021).

14. S. S. Chandran, Y. Wu, J. E. Subotnik, Effect of Duschinskii Rotations on Spin-Dependent Electron Transfer Dynamics, *J. Phys. Chem. A* **126**, 9535-9552 (2022)

15. T. J. Penfold, E. Gindensperger, C. Daniel, C. M. Marian, Spin-Vibronic Mechanism for Intersystem Crossing, *Chem. Rev.*, **118,** 6975–7025 (2018).

16. H. Behar-Levy, O. Neumann, R. Naaman, D. Avnir, Chirality Induction in Bulk Gold and Silver, *Adv. Mater*., **19**, 1207-1211 (2007).

17. W. Ma, L. Xu, A.F. De Moura, X. Wu, H. Kuang, C. Xu, N.A. Kotov, Chiral Inorganic Nanostructures, *Chem. Rev*., **117**, 8041-8093 (2017).

18. L. J. Van der Pauw, A method of measuring the resistivity and Hall coefficient on lamellae of arbitrary shape, *Philips Technical Review*. **20**, 220–224 (1958).

19. C. Seibel, J. H. Soh, S. Zilberg, and A. I. Krylov, Correction to: Ultrafast transversal CISS effect observed in a chiral photoswitching molecule, *J. Phys. Chem. Lett.* **16**, 9496 (2025).

20. T. A. Isaev and R. Berger, Polyatomic candidates for cooling of molecules with lasers from simple theoretical concepts, *Phys. Rev. Lett.* **116**, 063006 (2016).

21. J. F. Stanton and R. J. Bartlett, The equation of motion coupled-cluster method. a systematic biorthogonal approach to molecular excitation energies, transition probabilities, and excited state properties, *J. Chem. Phys.* **98**, 7029 (1993).

22. A. I. Krylov, Equation-of-motion coupled-cluster methods for open-shell and electronically excited species: The hitchhiker's guide to Fock space, *Ann. Rev. Phys. Chem*. **59**, 433 – 462 (2008)

23. H. A. Bethe and E. E. Salpeter, Quantum Mechanics of One and Two Electron Atoms (Plenum, New York, (1977).

24. P. Pokhilko, E. Epifanovsky, and A. I. Krylov, General framework for calculating spin–orbit couplings using spinless one-particle density matrices: theory and application to the equation-of-motion coupled-cluster wave functions, *J. Chem. Phys*. **151**, 034106 (2019).

25. B. P. Bloom, Y. Lu, T. Metzger, S. Yochelis, Y. Paltiel, C. Fontanesi, S. Mishra, F. Tassinari,R. Naaman. D. H. Waldeck, Asymmetric reactions induced by electron spin polarization, *Phys. Chem. Chem. Phys*., **22**, 21570 (2020).

26. X.Wang, M. Peralta, X. Li, P.V. Möllers, D. Zhou, P. Merz, U. Burkhardt, H. Borrmann, I. Robredo,C. Shekhar, H. Zacharias, X. Feng, C. Felser, Direct control of electron spin at an intrinsically chiral surface for highly efficient oxygen reduction reaction, PNAS 122, e2413609122 (2025)., Direct control of electron spin at an intrinsically chiral surface for highly efficient oxygen reduction reaction, *PNAS* **122**, e2413609122 (2025).

27. K. Banerjee-Ghosh, O. Ben Dor, F. Tassinari, E. Capua, S. Yochelis, A. Capua, S.-H. Yang, S. S. P. Parkin, S. Sarkar, L. Kronik, L. T. Baczewski, R. Naaman, Y. Paltiel, Separation of Enantiomers by Their Enantiospecific Interaction with Achiral Magnetic Substrates, *Science* **360**, 1331–1334 (2018).

28. S. F. Ozturk, Z. Liu, J. D. Sutherland, D. D. Sasselov, Origin of biological homochirality by crystallization of an RNA precursor on a magnetic surface, *Sci. Adv*. **9**, eadg8274 (2023).

29. S. F. Ozturk, D. K. Bhowmick, Y. Kapon, Y. Sang, A. Kumar, Y. Paltiel, R Naaman, D. D. Sasselov, *Nat. Comm*. **14**, 6351 (2023).

30. S. Furkan Ozturk, D. D. Sasselov, J. D. Sutherland, The central dogma of biological homochirality: How does chiral information propagate in a prebiotic network? *J. Chem. Phys.* **159**, 061102 (2023).

31. S. Miwa et al. Spin polarization driven by molecular vibrations leads to enantioselectivity in chiral molecules. *Sci. Adv*. **11**, eadv5220 (2025).
DOI: 10.1126/sciadv.adv5220.

32. P. Hedegård, Spin dynamics and chirality induced spin selectivity, *J. Chem. Phys.* **159**, 104104 (2023).

33. F. Evers et al, Theory of Chirality Induced Spin Selectivity: Progress and Challenges, Adv. Mater 34, 2106629, (2022). https://doi.org/10.1002/adma.202106629

34. J. Gersten, K. Kaasbjerg, A. Nitzan Induced spin filtering in electron transmission through chiral molecular layers adsorbed on metals with strong spin-orbit coupling. *J Chem Phys.* **139**, 114111 (2013).

35. L. Savi, L. Celada, D.K. A. P. Huu, A. Chiesa, S. Carretta, A. Painelli, Chirality-Induced Spin Selectivity: A Minimal Model, : J. Phys. Chem. Lett.,**16**, 9107–9115 (2025).
36. L. F. Chibotaru, Theoretical Understanding of Anisotropy in Molecular Nanomagnets, pages 185–229. Springer Berlin Heidelberg, Berlin, Heidelberg, 2015.
37. P.-˚A. Malmqvist and B. O. Roos, The CASSCF state interaction method, *Chem. Phys. Lett.* **155**, 189 (1989).
38. P.-˚A. Malmqvist, B. O. Roos, and B. Schimmelpfennig, The restricted active space (RAS) state interaction approach with spin–orbit coupling, *Chem. Phys. Lett.* **357**, 230 (2002).
39. M. Nooijen and R. J. Bartlett, Equation of motion coupled cluster method for electron attachment, *J. Chem. Phys*. **102**, 3629 (1995).
40. A. I. Krylov, Equation-of-motion coupled-cluster methods for open-shell and electronically excited species: The hitchhiker's guide to Fock space, *Annu. Rev. Phys. Chem*. **59**, 433 (2008).
41. J. H. Andersen, K. D. Nanda, A. I. Krylov, and S. Coriani, Probing molecular chirality of ground and electronically excited states in the UVvis and X-ray regimes: An EOM-CCSD study, *J. Chem. Theory Comput*. **18**, 1748 (2022).

## Figures

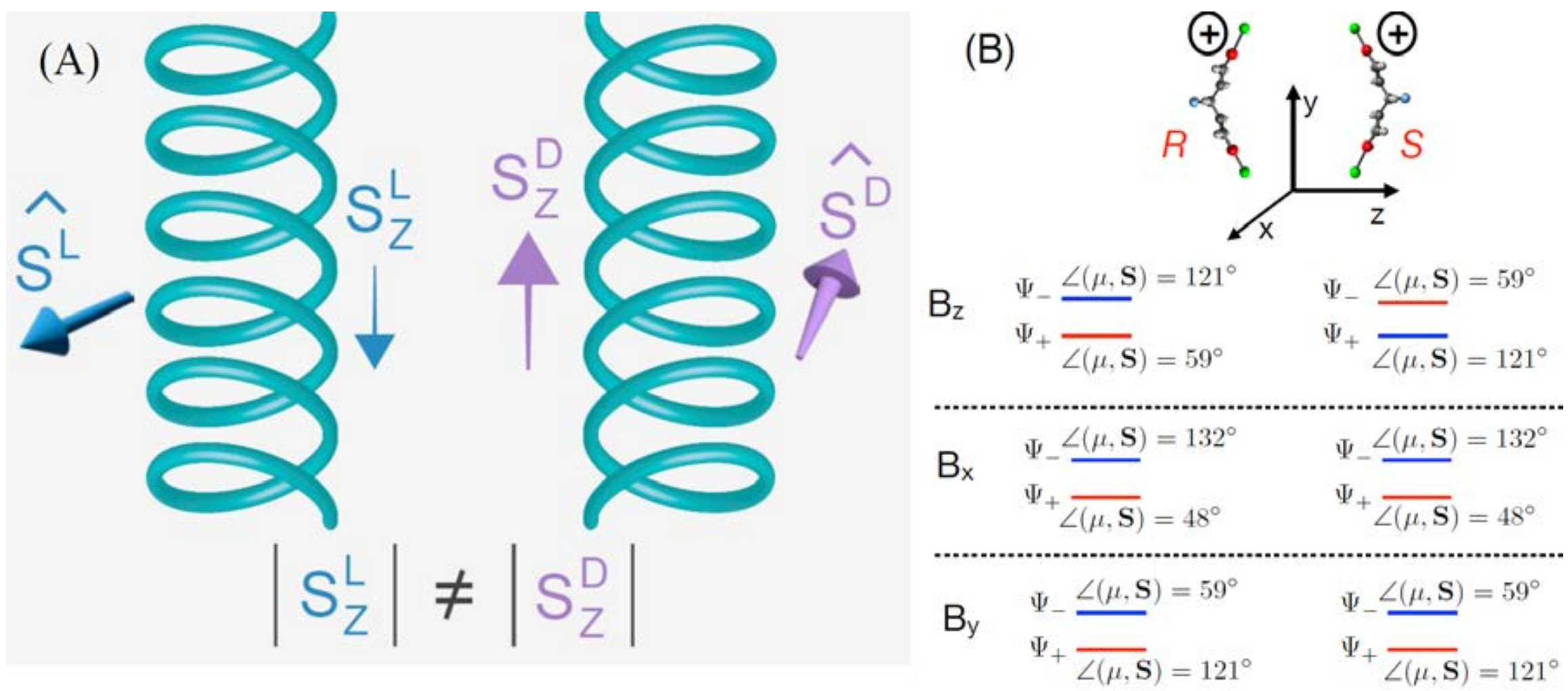

**Figure 1: The alignment of the total spin and the magnetic vectors. A)** A scheme of the spin momenta of the two enantiomers of a chiral system, which are not necessarily helical. The two total spin vectors, $\hat{S}^L$ and $\hat{S}^D$, are of the same magnitude but they are oriented at different angles relative to the long axis of the molecules (the angles add up to 180, as dictated by the time-reversal symmetry). As a result, their projection on the *z*-axis, $S_z$, differs for the two enantiomers. **B**) A scheme related to the system calculated by *ab initio* (Figure 3). State mapping in *R* and *S* enantiomers of the molecule studied in Ref. 8 (distrontium mono(1*E*,4*E*,3*R/S*)-3-fluoropenta-1,4-diene-1,5-bis(olate)) radical) shown and the corresponding values of the angle between the electric dipole and the total **S** vector, $\angle(\mu, S)$, for different choices of the quantization axis (set by B).

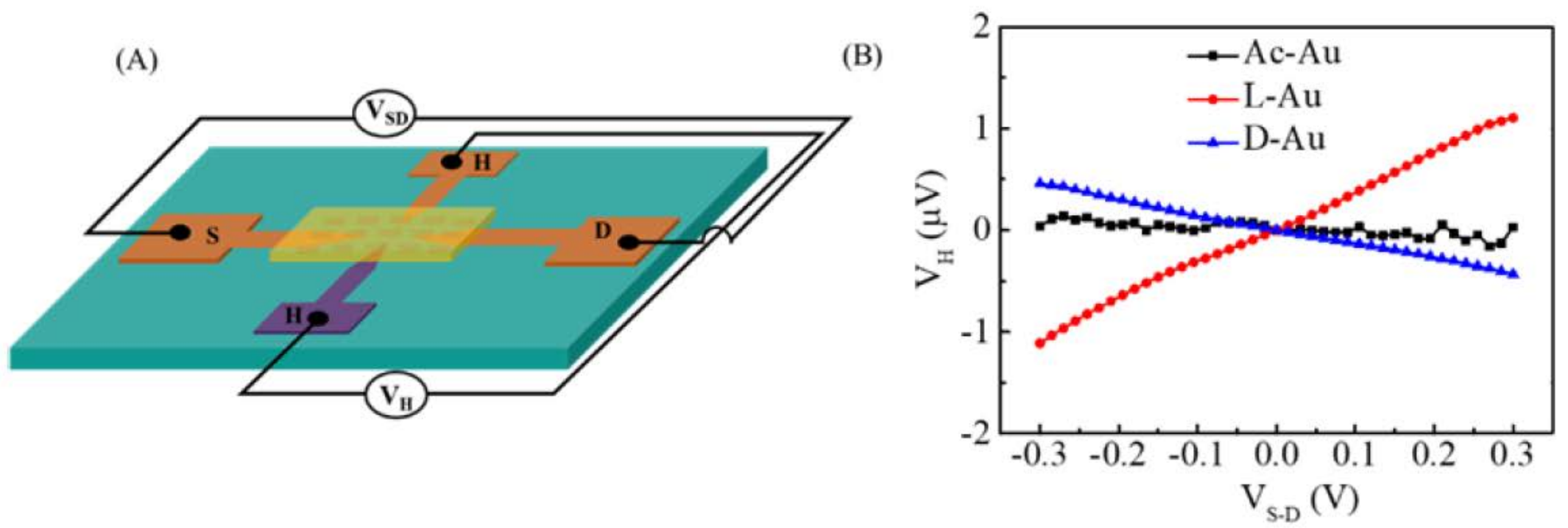

**Figure 2: The chirality-induced anomalous Hall signal**. (A) The Hall device used. (B) The Hall signals from the device where the distance between the source (S) and drain (D) electrodes is 4 µm. The Hall signal is dependent on the voltage applied on the gate. The slope of the Hall signal versus the applied gate voltage is positive for L-Au (red) and negative for D-Au (blue). In achiral gold (Ac-Au), there was no significant signal. (Copied with permission from ref. 5.)

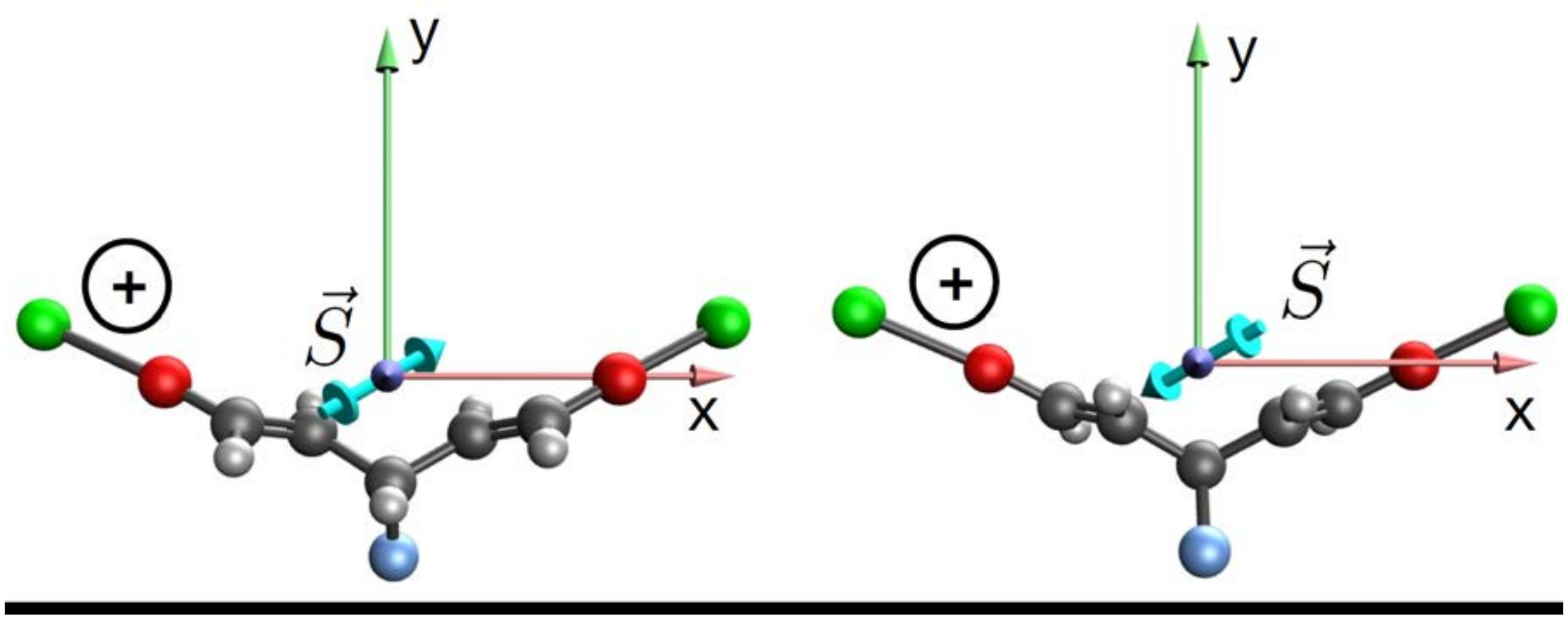

**Figure 3: Calculated total spin vector and its direction.** Spin-polarization illustrated by S (blue arrow) for a spin–orbit perturbed p-like excited state in the molecule studied in ref 8. A weak magnetic field along the x-axis is applied to resolve the states. Here the z-axis points at the viewer. Reproduced with permission from Ref. 22.

**Table 1:**

| | System | A% |
|---|---|---|
| 1. | Chiral gold | 28±6 |
| 2. | Chiral silver | 12±4 |
| 3. | Polyalanine adsorbed on gold | 34±10 |
| 4. | Polyalanine on gold + 3 nm oxide | 12±5 |

**Table 1:** The asymmetry, A, observed in the Hall signal for several molecules and configurations measured on 6 different devices for each of the four systems. Whereas the chiral gold and silver were prepared as described in the text, the L and D polyalanine molecules are a thiolated α-helix oligopeptide-based oligomer ((H]-C(AAAAK)$_7$-[OH]). The polylanine molecules were adsorbed on top of a Hall device either through a thiol group on a gold surface or with a carboxylic group on top of a 3 nm aluminum oxide layer deposited on the gold. The asymmetry factor is $A = \frac{S_L - S_D}{S_L + S_D}$ when $S_L$ and $S_D$ are the Hall signals observed for the L and D enantiomers, respectively.